\documentclass{PoS}
\usepackage{graphicx}
\usepackage{amsmath}
\usepackage{caption}

\title{FRG Approach to Nuclear Matter at Extreme Conditions}

\ShortTitle{FRG Approach to Nuclear Matter at Extreme Conditions}

\author{\speaker{P\'eter P\'osfay}\\
        Wigner Research Centre for Physics of the H.A.S., 29-33 Konkoly-Thege Mikl\'os Str., H-1121, Budapest, Hungary\\
        E\"otv\"os Lor\'and University, 1/A P\'azm\'any P\'eter Walkway, H-1117, Budapest, Hungary\\
        E-mail: \email{posfay.peter@wigner.mta.hu}}

\author{Gergely G\'abor Barnaf\"oldi\\
        Wigner Research Centre for Physics of the H.A.S., 29-33 Konkoly-Thege Mikl\'os Str., H-1121, Budapest, Hungary\\
        E-mail: \email{barnafoldi.gergely@wigner.mta.hu}}
        
\author{Antal Jakov\'ac\\
        E\"otv\"os Lor\'and University, 1/A. P\'azm\'any P\'eter Walkway, H-1117, Budapest, Hungary\\
        E-mail: \email{jakovac@caesar.elte.hu}}

\abstract{Functional renormalization group (FRG) is an exact method for taking 
into account the effect of quantum fluctuations in the effective action of the
system. The FRG method applied to effective theories of nuclear matter yields
equation of state which incorporates quantum fluctuations of the fields.
Using the local potential approximation (LPA) the equation of state for Walecka-type
models of nuclear matter under extreme conditions could be determined. These models
can be tested by solving the corresponding Tolman\,--\,Oppenheimer\,--\,Volkov (TOV) equations
and investigating the properties (mass and radius) of the corresponding compact 
star models. Here, we present the first steps on this way, we obtained a Maxwell construction within the FRG-based framework using a Walecka-type Lagrangian.}

\FullConference{The European Physical Society Conference on High Energy Physics\\
		22--29 July 2015\\
		Vienna, Austria}

\begin{document}

\section{Introduction}

Effective field theory can be used to describe the strongly interacting baryonic media well. Even more, the simplest Walecka-type model provides a good description for the ordinary nuclear matter. However, at low temperature and at high density limit one has to take into account quantum fluctuations too -- which corrections are usually neglected.  Especially, this is the case for the extreme dense matter of the cold compact celestial objects, the endpoints of the stellar evolution, such as like neutron, hybrid, or quark stars as well as white dwarfs.  

Our recent progress aims to explore the possibility of this improved description using the {\it functional renormalization group} (FRG) method to carry out the corresponding effective potential, including the above mentioned quantum fluctuations.

\section{An introduction to the FRG method}

The functional renormalization group method is a general way to find the effective action of a system. In the framework of FRG it is possible to calculate low energy effective (observable) quantities  by gradual momentum integration of a theory defined at some high energy scale, $k$. These low-scale effective quantities incorporate quantum fluctuations too. Using this method at finite temperature, it is possible to calculate the equation of state of the system, which contain quantum fluctuations. 

Technically using the FRG method led us to calculate the a quantum $n$-point correlation function by gradual path integration. The basic idea is to achieve this via introducing a regulator term, $R_{k,a b}$ in the generator functional, $Z_{k}[J]$, which acts as a mass term and suppress modes below scale, $k$ as explained in Refs.~\cite{Wetterich:1989xg,Gies:2006wv}
\begin{equation} \label{eq1}
Z_{k}[J]=\int \left( \prod_{a} d \psi_{a}\right) e^{-S[\psi]-\frac{1}{2}R_{k,a b} \psi_{a} \psi_{b} +\psi_{a} J_{a} }
\end{equation}
The scale-dependent effective action is the generator of the Feynman diagrams. This can be obtained by the Legendre-transformation of the \textit{Schwinger-functional}, using the usual definition,  $ W[J]=-i\ln{Z[J]}$, 
\begin{equation} \label{eq2}
\Gamma_{k}[J]=\underset{J}{\rm sup}\left( \psi_{a}J_{a} - W[J] \right) -\frac{1}{2}R_{k,a b} \psi_{a} \psi_{b},
\end{equation}
where $\psi_{a}=\delta \Gamma_{k}/ \delta J_{a}$. By introducing the regulator term, the effective action becomes scale-dependent and its scale dependence is governed by the Wetterich-equation~\cite{Wetterich:1992yh} 
\begin{equation} \label{eq3}
\partial_{k} \Gamma_{k}=\frac{1}{2} \,  {\rm STr} \, \left [ \left( \partial_{k}R_{k} \right) \left( \Gamma_{k}^{(2)} + R_{k} \right)^{-1} \right],
\end{equation}
where $\Gamma_{k}^{(2)}$ is the second derivative matrix of the effective action. This equation here, is given in the multi-index formalism where the indices run through momentum, position, flavour etc. The term 'STr' is stand for the normal \textit{trace} operation but includes a negative sign for fermionic fields. The low-scale (observable) effective action is computed by integrating the Wetterich-equation~\eqref{eq3}, from the classical limit,  at some UV-scale $k=\Lambda$ to the IR-scale $k=0$, where quantum effects are taken into account.  The initial condition in this integration is the UV-scale action $\Gamma_{k=\Lambda}$, which has to be chosen in a way, that the low-scale effective action reproduces physical quantities correctly. 

\section{The local potential approximation}

\begin{figure}[h]
\includegraphics[width=0.8\textwidth]{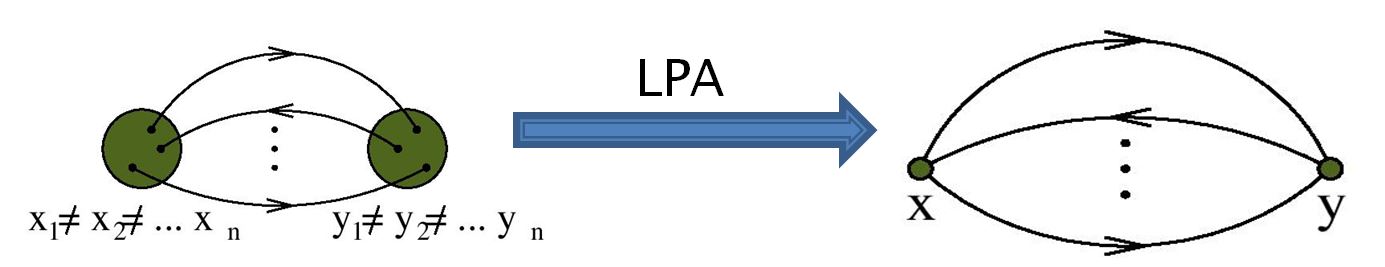}
\centering
\caption{The effect of the local potential approximation: at a certain scale the vertex appears as a point ({\it left}), below this
scale we can think of it as a patch of very close points ({\it right}). \label{fig:lpa}}
\end{figure}

The Wetterich-equation~\eqref{eq3} is exact but the integration is not feasible in most of the cases, because one has to include all possible operators in the effective action, even the most exotic ones, because they might become relevant at lower-scale values. Since this task is 
a challenging one, for practical purposes, a kind of truncation use to be introduced as an {\it ansatz} for the effective action. The \textit{local potential approximation} (LPA) is based on the assumption that propagators vary in spacetime much slower than vertices. This means that, the fields have different coordinate variables, but it is a good approximation to take their value at a common coordinate, which characterizes the vertex -- see Figure~\ref{fig:lpa}. This implies that the UV-scale effective action has the following form,
\begin{equation} \label{eq4}
\Gamma_{k}[\psi]= \int d^4 x \left[ \frac{1}{2}\psi_{i}K_{k,i j}\psi_{j}+ U_{k}\left(\psi \right) \right],
\end{equation} 
where $K_{k,i j}$ is the {\it kinetic kernel} and $U_{k}(\psi)$ is \textit{the scale-dependent effective potential}, which is some function (and not functional) of the fields. 

\section{FRG at finite temperature}

At finite temperature the path-integral~\eqref{eq1} extends to the imaginary time axes. Since the regulator-term is time-independent, the Kubo\,--\,Martin\,--\,Schwinger (KMS) relation~\cite{Lebellac} can be used to relate the propagator term appearing in equation~\eqref{eq3} to the spectral function,  $\rho(\omega)$ of the system,
\begin{equation} \label{eq5}
i\, G(\omega)=\left[ 1 + \alpha n_{\alpha}(\omega) \right]\rho (\omega), 
\end{equation}
where $n_{\alpha}(\omega)$ is the Fermi\,--\,Dirac or Bose\,--\,Einstein distributions respectively with $\alpha=\pm 1 $,
\begin{equation} \label{eq6}
n_{\alpha}(\omega)=\frac{\alpha}{e^{\beta \omega}-\alpha} \  \  .
\end{equation} 
Using (\ref{eq5}) and the LPA ansatz given by (\ref{eq4}), the Wetterich-equation gives an equation for the effective potential in the finite temperature limit
\begin{equation} \label{eq7}
\partial_{k}U=-\frac{1}{2} \int 
\frac{d^4 p}{(2 \pi)^4} \partial_{k} R_{i j}(\textbf{p})
\left[
\frac{1}{2}+n_{\alpha_{i}}(p_{0})
\right]
\rho_{i j}(p).
\end{equation} 
 
\section{The semi-finite temperature approximation}

Our goal is to find effective theories of nuclear matter under extreme conditions in the limit of zero temperature and finite chemical potential. To explore the behaviour of FRG equations in these conditions, first we use the following simplistic model, 
\begin{equation} \label{eq8}
\Gamma_{k}=\bar{\psi}\left(
{\not} p -m-g_{\sigma}\sigma 
\right)\psi +
\frac{1}{2}\left(
\partial_{\mu}\sigma \partial^\mu \sigma 
\right)
-U_{k}(\sigma).
\end{equation}
Inserting~\eqref{eq8} into~\eqref{eq7} the differential equation for the effective potential becomes
\begin{equation}
\label{dku}
\partial_{k}U_{k}=\frac{k^4}{12\pi^2}\left[\frac{2n_{b}(\omega_{\sigma})+1}{\omega_{\sigma}}
-8\frac{1-n_{f}(\omega-\mu)-n_{f}(\omega+\mu)}{\omega}\right],
\end{equation} 
where we introduced the following variables  as usual,
\begin{equation}
\label{omega}
\omega=\sqrt{k^2+(g_{\sigma}\sigma)^2} \qquad {\rm and } \qquad
\omega_{\sigma}=\sqrt{k^2+\frac{\partial^2 U_{k}}{\partial \sigma^2}}.
\end{equation} 
We note, eqs.~\eqref{eq8}-\eqref{dku} are a two variable second order non-linear partial differential equation. Since at small temperatures the Fermi\,--\,Dirac distribution behaves like a step function this forbids the use of implicit discretization with iterative solution of the resulting non-linear equation. However, using explicit methods one can solve the differential equation starting from a discretized potential at the UV-scale.
We can solve the equations at various temperatures, in particular at $T=0$. Then we can make the following approximation:  
assuming that the running of the potential does not depend strongly on the temperature. This should be true at low $T$, so it is possible to approximate the finite-temperature running of the potential with the zero-temperature running of the potential on the right hand side of eq.~\eqref{dku}. In this case eq.~\eqref{dku} become a simple integral with parameters $\mu$ and $\beta$. 

\begin{figure}[h]
\includegraphics[width=0.6\textwidth]{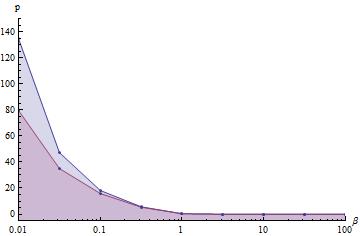}
\centering
\caption{Comparison of the pressure, $p$ as a function of $\beta$, for the case of the direct integration of the simplistic model  and in the semi-finite temperature approximation. (All are in scalar-mass units.)\label{semifinite} }
\end{figure}

The error of the above approximation can be estimated by solving eq.~\eqref{dku} explicitly at finite temperature and compare the solution to the integration based on the zero temperature running. The comparison of the two methods is shown on Figure~\ref{semifinite}. The two solutions for the pressure are very close for large  $\beta$ values down to $\beta\approx0.1$. This means that, in the low-temperature limit, it is a good approximation to use the integral of the zero temperature running.  This is especially true for compact star equation of state, where the typical temperature is usually orders of magnitude smaller than the binding energy per nucleon. Note, on Figure~\ref{semifinite} all the variables are measured in units of the scalar mass. 

\section{A Walecka-type model within the FRG method}

\begin{figure}[h]
\includegraphics[width=0.8\textwidth]{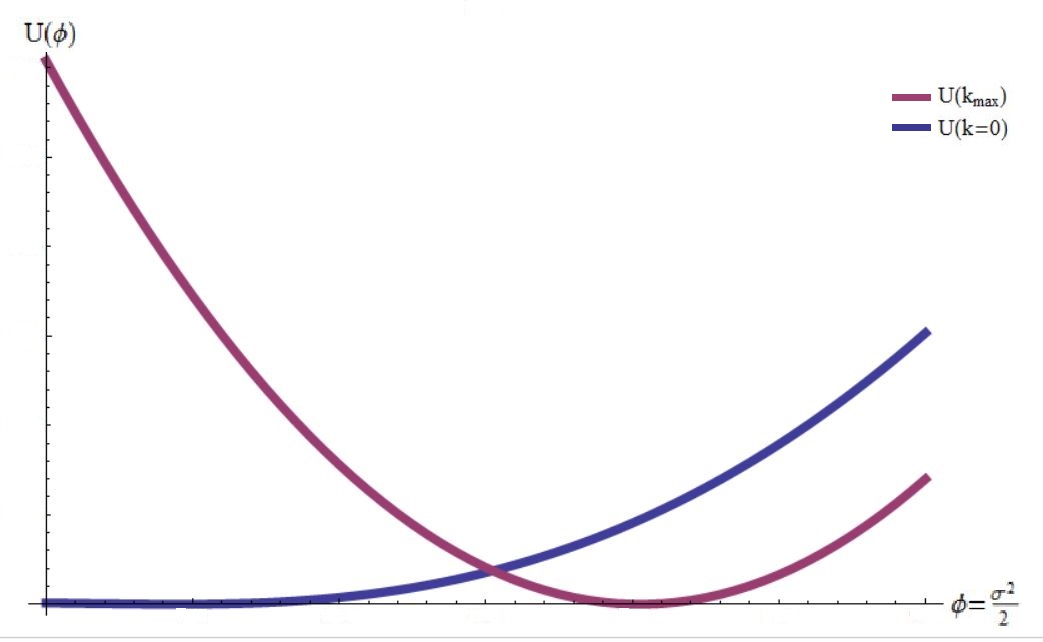}
\centering
\caption{{\it Color online:} The evolution of the potential, $U(\phi)$ in the Walecka-type model. The low-scale potential realizes the Maxwell-construction.  \label{max}}
\end{figure}

In order to test the validity of the functional renormalization group method,  a more realistic model was checked as well. Here, the evolution of the potential was studied in the context of an advanced model, characterized by the following Lagrangian 
\begin{equation}
\label{gammaW}
\begin{aligned}
\Gamma_{k}=
&\bar{\psi}\left[ 
\displaystyle{\not}p-g_{\sigma}(\sigma+i\gamma_{5}\tau_{j}\pi^j)-g_{\omega}\displaystyle{\not}\omega
\right]\psi \\
&+\frac{1}{2}\partial_{\mu}\sigma\partial^\mu\sigma
+\frac{1}{2}\partial_{\mu}\pi\partial^\mu \pi
-\frac{1}{4} F_{\mu\nu}F^\mu{}^\nu
+\frac{1}{2} m_{\omega}^2\omega_{\mu}\omega^\mu
-U(\sigma,\pi) \  .
\end{aligned}
\end{equation} 
This is very similar to the well-known Walecka-model~\cite{Walecka,glendenning}, but it contains an additional $\pi$ meson and a generalized potential for the $\sigma$ meson. We treat $\omega$ meson in mean field approximation as explained in Ref.~\cite{Drews:2014fya}. 
Following the receipt, the Wetterich-equation for the effective potential is given in the following form using the same notations and symbols as in eq.~\eqref{dku}:
\begin{equation}
\label{dkuW}
\partial_{k}U_{k}=\frac{k^4}{12\pi^2}\left[\frac{2n_{b}(\omega_{\sigma})+1}{\omega_{\sigma}}+3\frac{2n_{b}(\omega_{\pi})+1}{\omega_{\pi}}-8\frac{1-n_{f}(\omega-\bar\mu)-n_{f}(\omega+\bar\mu)}{\omega}\right]  , 
\end{equation} 
where
\begin{equation}
\omega_{\pi}=\sqrt{k^2+\frac{\partial U_{k}}{\partial \sigma}} \qquad  {\rm and} \qquad \bar\mu=\mu-g_{\omega}\omega_0.
\end{equation}
The integration of eq.~\eqref{dkuW} yields the potential, which is shown in Figure~\ref{max}. Here, the classical UV-scale potential ($U(k_{max})$, with {\it blue line}) and at  the quantum limit, with ($U(k=0)$,  with {\it red line}) was given in the following form:
\begin{equation}
\label{pot}
U(\phi)=-m^2 \phi + \lambda \phi^2, \qquad  {\rm where} \qquad \phi=\frac{\sigma^2}{2} \  .
\end{equation} 
The parameters for the Walecka-type model are $m^2=1.2 $ GeV\textsuperscript{2}, $\lambda=7.4$ and the cutoff scale is $\Lambda=1.3$ GeV. These parameters were chosen to reproduce the correct nucleon mass. During the evolution of the potential the expectation value of the $\sigma$ field is decreasing.  The smaller the value of the $k$, the potential flattens out as well in the regions where the curvature mass would be negative in the so called {\it coarse grained potential}. This can be understood that the potential realizes the Maxwell-construction. 
%

\section{Conclusions}

We have investigated a Walecka-type model by the functional renormalization group (FRG) method. We have calculated the running of the effective potential in local potential approximation (LPA) at zero and at finite temperature values assuming finite chemical potential. In the zero temperature limit we were able to reproduce the nuclear masses. This result supports the idea that the semi-finite temperature approximation works well in the low-temperature case. Finally, we obtained a Maxwell construction within the FRG-based framework using a Walecka-type Lagrangian. This achievement led us to our next step: to study the equation of state of the extreme dense nuclear matter exists e.g. in compact stars.

\section*{Acknowledgements}

This work was supported by Hungarian OTKA grants, NK106119, K104260, K104292, TET 12 CN-1-2012-0016 and NewCompStar COST MP1304 action. Author G.G.B. also thanks the J\'anos Bolyai research Scholarship of the Hungarian Academy of Sciences. Author P. P.  acknowledges the support by the Wigner RCP of the H.A.S. 


\end{document}